\documentclass[sigconf, pbalance]{acmart}

\usepackage{graphicx}
\usepackage[caption=false]{subfig}
\usepackage{multirow}
\usepackage{enumitem}

\usepackage{algorithm}
\usepackage{caption}
\usepackage[justification=justified,singlelinecheck=false]{caption}

\AtBeginDocument{%
  }

\setcopyright{none}
\copyrightyear{2025}
\acmYear{2025}
\acmDOI{XXXXXXX.XXXXXXX}
\acmConference[SC '25]{The International Conference for High Performance Computing, Networking, Storage, and Analysis.}{November 16--21, 2025}{St. Louis, MO}

\begin{document}

\title{
Advancing Quantum Many-Body GW Calculations on Exascale Supercomputing Platforms
}

\renewcommand{\shortauthors}{Zhang et al.}

\begin{abstract}
  \textbf{ 
  Advanced \textit{ab initio} materials simulations face growing challenges as increasing systems and phenomena complexity requires higher accuracy, driving up computational demands. Quantum many-body GW methods are state-of-the-art for treating electronic excited states and couplings but often hindered due to the costly numerical complexity. Here, we present innovative implementations of advanced GW methods within the BerkeleyGW package, enabling large-scale simulations on Frontier and Aurora exascale platforms.
  Our approach demonstrates exceptional versatility for complex heterogeneous systems with up to 17,574 atoms, along with achieving true performance portability across GPU architectures. We demonstrate excellent strong and weak scaling to thousands of nodes, reaching double-precision core-kernel performance of 1.069 ExaFLOP/s on Frontier (9,408 nodes) and 707.52 PetaFLOP/s on Aurora (9,600 nodes), corresponding to 59.45\% and 48.79\% of peak, respectively.  Our work demonstrates a breakthrough in utilizing exascale computing for quantum materials simulations, delivering unprecedented predictive capabilities for rational designs of future quantum technologies.
 }
\end{abstract}




\begin{teaserfigure}
\vspace{-0.6cm}  
\begin{center}
    \large
    Benran Zhang\textsuperscript{1},
    Daniel Weinberg\textsuperscript{2},
    Chih-En Hsu\textsuperscript{1,3},
    Aaron R. Altman\textsuperscript{4},
    Yuming Shi\textsuperscript{4},
    James B. White III\textsuperscript{5},
    Derek Vigil-Fowler\textsuperscript{6}, \\
    Steven G. Louie\textsuperscript{2,7}, 
    Jack R. Deslippe\textsuperscript{2},
    Felipe H. da Jornada\textsuperscript{4,8}, 
    Zhenglu Li\textsuperscript{1,}*,
    Mauro Del Ben\textsuperscript{2,}* \\
    \normalsize
    \vspace{0.2cm}
    \textsuperscript{1}University of Southern California, USA.
    \textsuperscript{2}Lawrence Berkeley National Laboratory, USA.
    \textsuperscript{3}Tamkang University, Taiwan. \\
    \textsuperscript{4}Stanford University, USA.
    \textsuperscript{5}Oak Ridge National Laboratory, USA.
    \textsuperscript{6}National Renewable Energy Laboratory, USA. \\
    \textsuperscript{7}University of California at Berkeley, USA.
    \textsuperscript{8}SLAC National Accelerator Laboratory, USA. \\
    *Correspondence to be addressed to: zhenglul@usc.edu (Z.L.), mdelben@lbl.gov (M.D.B.)
\end{center}
\vspace{0.2cm}  
\end{teaserfigure}

\maketitle

\section{Justification for ACM Gordon Bell Prize}
BerkeleyGW delivers a breakthrough in exascale computational quantum many-body materials simulations, achieving true performance portability across all leadership-class supercomputing architectures, excellent strong scaling, and high fraction of peak for core computing kernels. 
BerkeleyGW enables predictive quantum materials modeling with outstanding time-to-solution and double-precision throughput above 1.0 ExaFLOP/s performance.

\begin{figure*}
    \centering
    \includegraphics[width=1.0\textwidth]{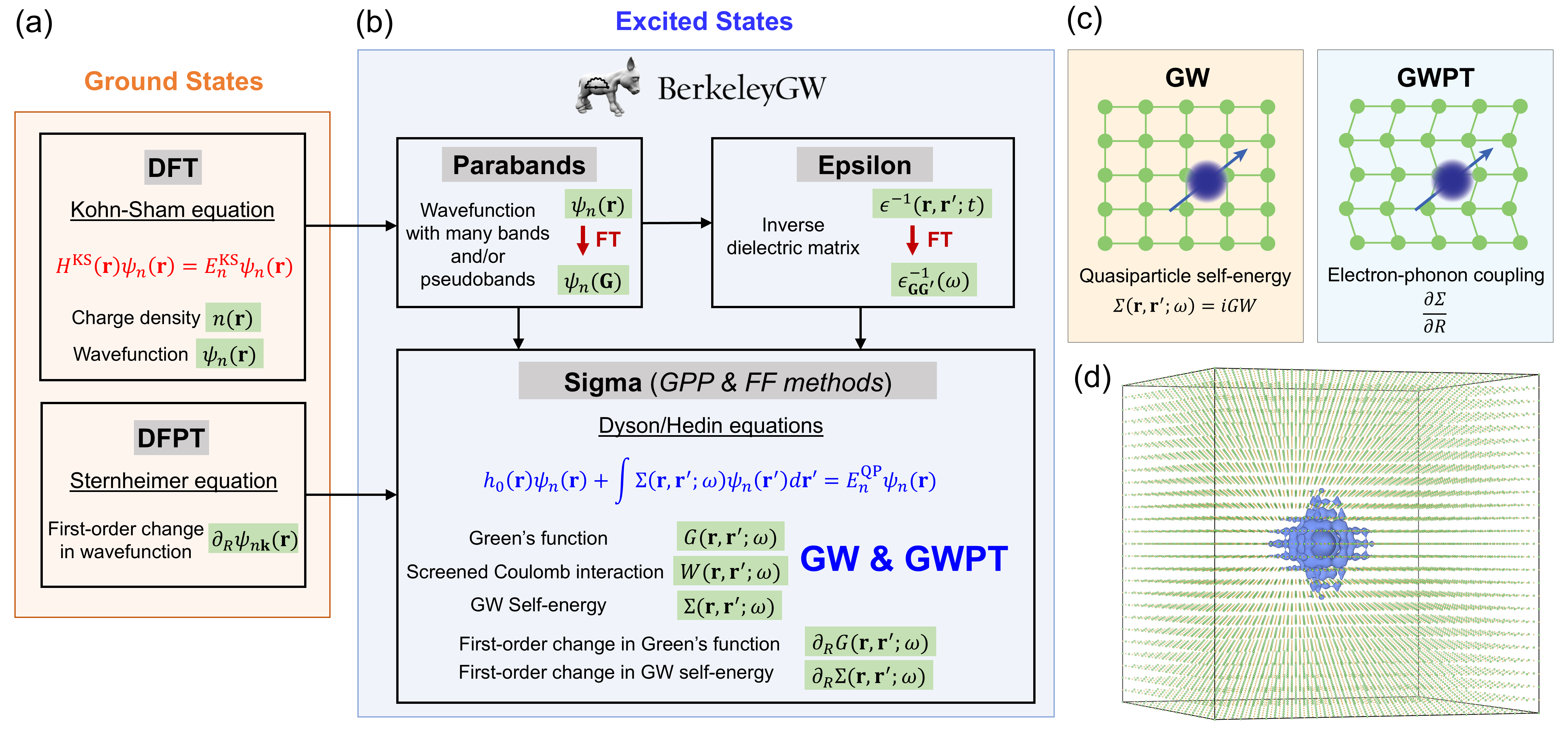}
    \caption{Workflow of GW and GWPT calculations. \textmd{(a) Ground-state DFT and DFPT calculations (starting-point for GW). (b) Excited-state calculations using BerkeleyGW. Various key quantities are highlighted in the green shade (FT = Fourier transform). 
    (c) The GW method computes quasiparticle excitation. The GWPT method computes electron-phonon coupling ($R$ represents atom positions). (d) Charge density of a defect state in the LiH17,574 system. }}
    \label{fig:overview}
\end{figure*}

\section{Performance Attributes}
\begin{table}[h!]
\vspace{-0.3cm}
 \begin{tabular}{|@{\;\,}l@{\;\,}|@{\;\,}l@{\;\,}|}
    \hline
    Category of Achievement & Scalability, Time-to-solution, \\
    & Peak Performance\\ \hline
    Type of Method Used & Both Explicit and Implicit \\ \hline
    Results Reported & Kernel Only \\
    \hline
    Precision Reported & Double Precision \\ \hline
    System Scale & Results Measured on Full System \\ \hline
    Measurement Mechanism & FLOP Count  \\
    \hline
 \end{tabular}
\end{table}

\section{Overview of the Problem}

Quantum materials research has entered a new era where a broad range of emerging materials systems and various many-body phenomena are becoming central topics of studies. This manifests in two aspects: first, the materials systems are becoming increasingly heterogeneous, such as defects in semiconductors as solid-state qubits (e.g., nitrogen-vacancy center) and nanometer-scale moir\'e superlattices (e.g., twisted bilayer graphene); second, the dominant interactions of interest often have quantum many-body nature, such as electron-electron, electron-phonon, and electron-hole interactions. Predictive first-principles, or \emph{ab initio} methods, are heavily demanded to understand these complex phenomena and systems and to design next-generation electronic, optical, and quantum devices.

Standard density functional theory (DFT) approaches nowadays are able to process heterogeneous systems of $O(10^4)$ of atoms, as achieved recently in the Gordon Bell Prize in 2023 \cite{10.1145/3581784.3627037}. However, DFT methods bear significant limitations in describing excited-state properties of materials, which requires explicit calculations of the electron-electron interactions that are not captured within DFT. DFT-based approaches have lagged, both quantitatively and even qualitatively, behind the desired predictive power from first principles for issues such as band gaps and electron-phonon coupling strengths -- quantities critical for the design of novel materials for energy and quantum science applications, for instance.

The GW approximation \cite{hybertsen1,hybertsen2,Hedin1965-lb} (where $G$ stands for Green's function and $W$ for the screened Coulomb interaction) is one approach for capturing the explicit electron-electron interactions in materials, and has been successful in predicting band gaps, band widths, and molecular excitation energy levels accurately. 
Moreover, extensions based on other first-principles formalisms building on top of the GW approximation have demonstrated excellent accuracy in many excited-state phenomena. 
For example, the first-principles GW plus Bethe-Salpeter equation approach \cite{onida2002electronic} can comprehensively describe optical spectra and excitonic properties of materials ranging from bulk solids to two-dimensional (2D) materials to molecules. More recently, the development of GW perturbation theory \cite{Li_GWPT_2019} (GWPT)  has enabled systematic electron-phonon coupling calculations at the many-electron level and shown excellent improvement against the results based on density-functional perturbation theory (DFPT) in several quantum materials \cite{Li_GWPT_2019,li2021unmasking}.

The BerkeleyGW software package offers a range of widely-adopted first-principles methodologies for electronic and optical excitations based on the GW approximation, as well as unique developments such as GWPT for correlated electron-phonon coupling. To apply these accurate quantum many-body methods to study complex quantum materials systems such as solid-state defects and moir\'e superlattices, the computational bottleneck of such approaches must be addressed. In 2020, BerkeleyGW was successfully scaled to the full machine of Summit with the NVIDIA GPU architecture \cite{DelBen2020_GB}.
In the last few years, the first-ever exascale supercomputers Frontier and Aurora became available with, however, different hardware architectures using AMD and Intel GPUs, respectively. It poses a compelling demand to port large-scale computational software packages to adapt to various GPU architectures (namely, AMD, Intel, and NVIDIA) while keeping the high performance.

The GW method computes electronic quasiparticle (QP) excitations in materials by solving the Dyson's equation,
\begin{equation}
\label{eq:dyson_equation}
h_0(\mathbf{r})\psi_{i}(\textbf{r}) 
+ \int \Sigma(\textbf{r}, \textbf{r}'; E_{i}^{\text{QP}})\psi_{i}(\textbf{r}') d\textbf{r}' =E_{i}^{\text{QP}}\psi_{i}(\textbf{r}) \ \ ,
\end{equation}
where \textbf{r} is the electron position, $\psi_i$ is the quasiparticle wavefunction of state $i$, $E_{i}^{\text{QP}}$ is the quasiparticle energy, $h_0$ is the mean-field Hamiltonian, and $\Sigma$ is the self-energy in the GW approximation derived from Hedin's equations \cite{Hedin1965-lb}. The self-energy operator describes the non-local and frequency-dependent quantum mechanical interactions among the electrons. The complexity of the GW method is much higher than the widely used DFT method, as shown in Fig. \ref{fig:overview} and can be easily seen in the number of arguments in corresponding operators, i.e., DFT: $(\textbf{r})$; GW: $(\textbf{r}, \textbf{r}'; \omega)$.

In solving Eq.~\ref{eq:dyson_equation}, the self-energy matrix elements $\Sigma_{lm}(E)$ can be constructed by a set of (many) $N_b$ wavefunctions $\{ \psi_n\}_{n=1..N_b}$ (also referred to as bands or states) \cite{hybertsen1,hybertsen2}:
\begin{equation}
\begin{aligned}
\label{eq:ff-ch}
\Sigma_{lm}(E) & =\frac{i}{2\pi}\sum_{nGG'}
{M_{l n}^{-G}}^{*} M_{m n}^{-G'}\int_0^\infty d\omega
\frac{ \epsilon^{-1}_{ GG'}\left(\omega \right)  v_{G'}}
{E-E_n-\omega}
\ \ ,
\end{aligned}
\end{equation}
where $l,m$ are quasiparticle band or state indices of interest, $n$ runs over the whole $N_b$ bands range, $G$ labels planewave (PW) basis elements,  $\epsilon^{-1}_{GG'}$ is the inverse dielectric matrix of the system in planewave basis, $E_n$ is the orbital energy, $v_{G'}$ is the Coulomb interaction in the reciprocal space, and $M$ represents the plane-wave matrix elements of wavefunctions,
$   M_{m n}^{G} = \int  d\mathbf{r}\; \psi^*_{m}(\mathbf{r})  e^{i\mathbf{G}\cdot\mathbf{r}}  \psi_{n}(\mathbf{r})$. 
The frequency ($\omega$) integration can be very well treated via the generalized plasmon-pole (GPP) model~\cite{hybertsen1}, as well as the direct full-frequency (FF) sampling.

The inverse epsilon matrix $\mathbf{\epsilon^{-1}}$ is constructed with the polarizability matrix $\chi_{GG'}$,
\begin{equation}
\label{eq:epsilon_mat}
\mathbf{\epsilon^{-1}}(\omega) = \left[ \mathbf{I - v \chi}(\omega) \right]^{-1} \ \ ,
\end{equation}
where $\mathbf{I}$ and $\mathbf{v}$ are diagonal identity and Coulomb matrices, and,
\begin{equation}
\label{eq:chi_mat}
 \chi_{GG'}(\omega) = 2 \sum_{v c} {M_{v c}^{G}}^{*} \Delta_{vc}(\omega) M_{v c}^{G'} \ \ .
\end{equation}
Here $v, c$ are wavefunction indices spanning the $N_v$ valence and $N_c$ conduction states. Note that $N_v+N_c=N_b$. $\Delta_{v c}(\omega)$ is an energy factor containing the orbital energies $E_v$, $E_c$,  and dependence on frequency $\omega$. 

A standard BerkeleyGW workflow (see Fig. \ref{fig:overview}) starts with the ground-state DFT (and DFPT) calculations to generate input for excited-state GW (and GWPT) calculations. Typically, many bands (up to thousands or tens of thousands) are needed for convergence: a challenge for iterative solvers in most DFT codes. BerkeleyGW provides a Parabands module that can generate a large set of wavefunctions $\{\psi_n\}$ of $N_b$ bands based on DFT output. The Epsilon module computes the inverse dielectric matrix. The Sigma module constructs the self-energy operator and evaluates a set of self-energy matrix elements ($N_\Sigma$ diagonal elements, or $N_\Sigma^2$ full matrix elements including off-diagonal ones), for the GW quasiparticle excitation energies $E^{\text{QP}}$ and  GWPT electron-phonon matrix elements. 
For practical implementations of Eqs.~\ref{eq:ff-ch}, \ref{eq:epsilon_mat} and \ref{eq:chi_mat}, the canonical GW method has an overall $O(N^4)$ scaling with standard calculation parameters summarized in Table~\ref{tab:comput_params}.

In this work, we present several significant methodological and algorithmic innovations implemented in BerkeleyGW: 
\begin{itemize}[left=0pt]
    \item True portability across AMD, Intel, and NVIDIA GPU architectures using \textit{directive-based} OpenACC and OpenMP models demonstrated on Frontier, Aurora, and Perlmutter.
    \item Kernel optimizations with \textit{hardware-optimized} programming languages, i.e., HIP for AMD, SYCL for Intel, and CUDA for NVIDIA GPUs, reaching high fraction of peak performance. 
    \item Excellent strong and weak scaling up to (nearly) the full machine of Aurora and Frontier.
    \item  
    New optimized kernels achieving high FP64 throughput of 1.069 EFLOP/s on Frontier (9,408 nodes or 75,264 GPUs) and 707.52 PFLOP/s on Aurora (9,600 nodes or 115,200 GPUs), corresponding to 59.45\% and 48.79\% of the theoretical and attainable peak, respectively.
    \item 
    Massive materials applications including silicon (Si) divacancy (up to 2,742 atoms)
    and lithium hydride (LiH) defect (up to 17,574 atoms).
    \item Advanced GW methods, including GWPT for electron-phonon coupling, full-frequency (FF) GW, and reduced scaling GW with mixed stochastic-deterministic approach.
\end{itemize}

\vspace{-1mm}
\begin{table}[ht!]
 \caption{Computational parameters in the GW workflow.}
\label{tab:comput_params}
\centering
\small 
\vspace{-0.3cm}
\begin{tabular}{ c c }
 \toprule
Symbol         &  Synopsis \tabularnewline
 \midrule
$N_{G}^{\psi}$ & No. of PWs ($G$ vectors) for wavefunctions  $\{ \psi_n\}$ \tabularnewline
$N_{G}$ & No. of PWs ($G$ vectors) for $\epsilon$, $\chi$ (Eq.~\ref{eq:epsilon_mat},\ref{eq:chi_mat}) \tabularnewline
$N_v$ & No. of valence bands (Eq.~\ref{eq:chi_mat}) \tabularnewline
$N_c$ & No. of conduction bands (Eq.~\ref{eq:chi_mat}) \tabularnewline
$N_b$ & No. of total bands $N_v+N_c$ (Eq.~\ref{eq:ff-ch}) \tabularnewline
$N_\Sigma$ & Dimension of $\Sigma(E)$ self-energy matrix (Eq.~\ref{eq:ff-ch}) \tabularnewline
$N_E$ & No. of $E$ grid points for $\Sigma(E)$ (Eq.~\ref{eq:ff-ch}) \tabularnewline
$N_{\omega}$ & No. of $\omega$ integration points (Eq.~\ref{eq:ff-ch})  \tabularnewline
$N_{\text{Eig}}$ & No. of eigenvectors for low rank $\chi^0(\omega)$  \tabularnewline
$N_p$ & No. of phonon perturbations $R_p$ (Eq.~\ref{eq:gwpt}) \tabularnewline
 \bottomrule
 \multicolumn{2}{@{}c}{All parameters grow linearly with system size except $N_E$ and $N_{\omega}$.}
\end{tabular}
\end{table}
\vspace{-2mm}

\section{Current State of the Art}
The GW approximation, originally derived by Hedin \cite{Hedin1965-lb}, has been a successful \emph{ab initio} approach to obtaining quasiparticle properties since the seminal work by Hybertsen and Louie \cite{hybertsen1,hybertsen2}. These properties allow for accurate understanding of energy levels and their alignments in a variety of environments, making GW the theory of choice for studying heterogeneous and extended systems. 

    
As described above, the traditional sum-over-states formulation of GW calculations has a formal scaling of $O(N^4)$, and different approaches have been taken to enhance the computational efficiency and scaling of GW calculations. In one approach, the summation over empty states is eliminated by using DFPT. This has been implemented successfully in multiple code bases by Umari \textit{et al.} \cite{umari}, Giustino \textit{et al.} \cite{giustino1}, and Govoni \textit{et al.} \cite{west1,west2,west3}. While the scaling of this approach remains $O(N^4)$, it has a distinct advantage in avoiding the generation of wavefunctions of many empty states.
Another approach that stays within the traditional sum-over-states paradigm is to use so-called pseudobands that are stochastic averages over the Kohn-Sham (KS) states within defined energy windows. Altman \textit{et al.}~\cite{Altman2024-mp} have shown that the use of pseudobands greatly reduces the number of bands needed to obtain accurate quasiparticle energies, and reduces the scaling of GW calculation to $O(N^{2.4})$. 
The pseudobands concept is inspired by the fully stochastic GW approach~\cite{PhysRevLett.113.076402}, which shows linear scaling for the computation of certain materials properties. 
However, stochastic GW introduces uncorrelated stochastic errors, and cannot compute all electronic properties available from deterministic GW calculations. 
Real-space, imaginary-time GW was originally proposed and implemented by Rieger \textit{et al.} \cite{Rieger1999-cl}, reducing the system size scaling to $O(N^3)$. Several implementations have used this scheme or related approaches \cite{PhysRevB.94.165109,doi:10.1021/acs.jpclett.7b02740,PhysRevB.101.035139}. However, this is achieved through various transformations that lead to a significant prefactor, so that the system size at which the $O(N^3)$ space-time GW method becomes favorable is highly system dependent. Yeh and Morales used interpolative density fitting to also achieve $O(N^3)$ scaling and argued that the prefactor should generally be smaller than that in the space-time approach \cite{Yeh2024-jt}. However, their calculations were done on downfolded model Hamiltonians, so the full promise of this approach for first-principles GW calculations remains to be determined. 

The GW method is often implemented within the planewave basis set, such as the popular software packages BerkeleyGW \cite{bgw} (this work), WEST \cite{west1,west_gpu}, Quantum ESPRESSO \cite{quantum-espresso}, Abinit \cite{abinit1}, Yambo \cite{yambo}, SternheimerGW \cite{sternheimergw}, and VASP \cite{vasp1,vasp2}. Other implementations of the GW method use localized basis functions, such as numerical atomic orbitals (FHI-aims~\cite{BLUM20092175}), Gaussians (Fiesta~\cite{doi:10.1021/acs.jctc.5b00304}, MolGW~\cite{BRUNEVAL2016149}), linearized augmented-planewave with local orbitals (Exciting~\cite{gulans2014exciting}, ELK~\cite{ELK_code}), and mixed Gaussian and planewaves (CP2K~\cite{doi:10.1063/5.0007045}). Localized basis sets typically have reduced computational cost due to their smaller basis size, but convergence has to be checked carefully in systems with diffused states. Generally, planewave implementations are more suitable for extended systems, while localized basis sets are used for localized systems such as molecules and nanoclusters. 


As most GW studies still focus on systems with tens to hundreds of atoms because of the high computational complexity, the community is pushing towards much larger systems to access new phenomena.
Among the largest GW calculations to date, we mention here (i) the twisted bilayer phosphorene structure containing $\sim$2,700 atoms ($\sim$13.5k electrons) using linear-scaling stochastic GW by Brooks \textit{et al.}~\cite{stoGW-morie}, (ii) our previously reported result of silicon divacancy defect with 2742 atoms ($\sim$11k electrons) using the planewave basis set and the deterministic approach with BerkeleyGW~\cite{DelBen2020_GB}, (iii) the $\sim$10k electrons calculation with WEST by Yu \textit{et al.}~\cite{west_gpu}, and (iv) the recent result of $\sim$14k atoms/electrons calculation of LiH using a low-rank approximation approach by Wu \textit{et al.}~\cite{10.1109/SC41406.2024.00067}. 
In this work, we focus on optimizing the standard GW approach with planewave basis sets as implemented in BerkeleyGW~\cite{bgw}.  Our results achieve unprecedented scalability and applicability to describe complex materials on exascale platforms.

\section{Innovations Realized}

We present recent theoretical, algorithmic, and HPC optimization advances in BerkeleyGW. Sec.~\ref{sec:gwpt} introduces the first-of-its-kind GW perturbation theory (GWPT) to study correlated electron-phonon coupling. Sec.~\ref{sec:gwff} details methods to accelerate full-frequency GW calculation by reducing its $O(N_G^2)$ cost dependence and $O(N^3)$ memory bottleneck. Sec.~\ref{sec:pseudob} addresses the need for large $N_b$ via a mixed stochastic-deterministic scheme that compresses the wavefunction space and effectively lowers the $O(N^4)$ scaling. Sec.~\ref{sec:portability} outlines performance portability across pre-exascale and exascale platforms. Sec.~\ref{sec:kernel_opt_diag} and Sec.~\ref{sec:fullmat_kernel_opt} report GPU kernel optimizations for diagonal and off-diagonal self-energy matrix elements, pushing performance to high peak and enabling large-scale GW and GWPT calculations, including full solutions to Dyson's equation.

\subsection{GW Perturbation Theory}\label{sec:gwpt}

Electron-phonon coupling is one of the central interactions in materials physics, and is critical to a wide range of materials properties, including carrier mobility, optical absorption, quantum decoherence, and phonon-mediated superconductivity, among others. Accurate first-principles computation of microscopic electron-phonon interactions of materials systems is essential to design and optimize next-generation electronic and optoelectronic devices. The prevailing approach for systematic calculations of electron-phonon matrix elements is DFPT, which is a linear-response theory of DFT. The linear-response formulation elegantly decomposes the phonon perturbations to an electronic system into independent modes, where each perturbation can be solved at a similar cost as a standard DFT calculation. However, DFPT inherits similar limitations as DFT, and becomes insufficient for materials with stronger electron correlation effects.

GWPT is a newly developed method that enables systematic computation of electron-phonon coupling at the many-body level within the linear-response formulation for the first time \cite{Li_GWPT_2019}. GWPT has demonstrated excellent accuracy in capturing the correlation effects in the electron-phonon coupling beyond DFPT in several quantum materials. In GWPT, the equation to compute the atom-displacement-perturbed self-energy operator matrix elements is,
\begin{equation} \label{eq:gwpt}
\begin{split}
       \left[ \frac{\partial}{\partial R_p}\Sigma(E) \right]_{lm} = & \frac{i}{2\pi}\sum_{nGG'}
\biggr[ \frac{\partial {M_{l n}^{-G}}^{*}}{\partial R_p}  M_{m n}^{-G'} + {M_{l n}^{-G}}^{*} \frac{\partial M_{m n}^{-G'}}{\partial R_p}   \biggr] \\
        & \times \int_0^\infty d\omega
\frac{ \epsilon^{-1}_{ GG'}\left(\omega \right)  v_{G'}}
{E-E_n-\omega} \ \ ,
\end{split}
\end{equation}
where the operator $\frac{\partial}{\partial R_p}$ represents an atom-displacement induced perturbation, where $p$ labels the degrees of freedom (e.g., a particular atom moving along one direction, or a phonon eigenmode). The construction of the first-order change in self-energy $\partial_R \Sigma$ needs the first-order changes in the matrices $\partial_R M$, which are constructed by first-order changes in the wavefunctions $\partial_R \psi_{n}$ of all $N_b$ bands.

BerkeleyGW is the only package offering the implementation of the unique GWPT method. Currently, the frequency dependence is treated within the GPP model, which provides a straightforward strategy for implementation. Moreover, the GPP kernel has been extensively optimized for excellent peak performance, alleviating the high computational burden of GWPT that introduces an additional prefactor $N_p$ (number of perturbations) to the complexity of the standard GW. 
On the other hand, the $N_p$ perturbations are independent and massively parallelized to full scale with minimal communications on exascale machines.

\subsection{Fast Full-Frequency GW Method}\label{sec:gwff}



Although the treatment of the frequency dependence of the polarizability $\chi(\omega)$ has often been achieved via the GPP model, recent advances have enabled direct calculations of the full-frequency dependence, at a competitive computational cost. The key advance is the static subspace approximation~\cite{DelBen2019b}, where the zero-frequency polarizability $\chi(\omega=0)$ is firstly calculated (using Eq. \ref{eq:chi_mat}), then a subspace is defined with $\chi(\omega=0)$  to calculate non-zero frequencies~\cite{west1,west_wilson2008,west_wilson2009,west2,west3}. The zero-frequency polarizability is diagonalized and the $N_\text{Eig}$ most significant eigenvectors are kept as the subspace basis. The non-zero-frequencies polarizability can then be constructed as a modified Eq. \ref{eq:chi_mat},
\begin{equation}
\label{eq:chi_mat_sub}
 \chi_{BB'}(\omega\neq0) = 2 \sum_{v c} {M_{v c}^{B}}^{*} \Delta_{vc}(\omega) M_{v c}^{B'} \ \ ,
\end{equation}
where $B$ and $B'$ index over $N_\text{Eig}$ subspace basis vectors. The subspace representations of $M_{vc}$ and $\chi$ are connected to their planewave representations via the $N_G \times N_\text{Eig}$ projection matrix $\mathbf{C}_s$, ${M_{v c}^{B}}  = \sum_G M_{vc}^G C_s^{GB}$ and $\chi_{BB'} = \sum_{GG'} (C_s^{GB})^*\chi_{GG'} C_s^{G'B'}$.
This subspace compression reduces the computation of $\chi(\omega)$ from $O(N_\omega N_v N_c N_G^2)$ to $O(N_v N_c N_G^2+ N_\omega N_v N_c N_\text{Eig}^2)$ since the full planewave basis is only used for the zero frequency. In general, a subspace fraction of 10-20\% is sufficient to converge GW quasiparticle energies, hence this approximation results in a speedup of $\sim25-100$ times over the full planewave implementation~\cite{DelBen2019b,rpa-application}.

Full-frequency polarizability calculations are further enabled by careful GPU offloading of the key computational kernels. Significant prior efforts have been spent in optimizing the Fourier transformations to obtain the planewave matrix elements $M^G_{nm}$ (\textit{MTXEL} kernel~\cite{DelBen2020_GB}). Calculation of the polarizability via Eq. \ref{eq:chi_mat_sub} in  \textit{CHI\_SUM} kernel is most computationally intensive, which suffers from an $O(N^3)$ memory footprint on both host and device. 
To address this issue, our redesigned implementation effectively divides the computation into workable blocks over the $N_v$ valence bands, which we call the NV-Block algorithm. Combination of NV-Block and static subspace approximation enables efficient and accurate calculation of the full-frequency polarizability~\cite{rpa-implementation}. 
Full-frequency self-energy calculations also benefit from the subspace approximation by performing the $G$ and $G'$ sums in the reduced basis set (Eq. \ref{eq:ff-ch}), where the key steps can be casted as dense matrix multiplication.

The full-frequency GW offloading was solely performed using the open programming models OpenMP-target and OpenACC, which enables portability across the various leadership HPC systems and reduces the development overhead. Since most of the computationally limiting kernels could be offloaded to vendor matrix multiplication libraries, the use of the open models was less of a hindrance to performance. 

\subsection{Reduced Cost and Scaling with a Mixed Stochastic-Deterministic Algorithm}\label{sec:pseudob}

A major bottleneck of GW calculations is the sum-over-bands in inverse dielectric matrix $\epsilon^{-1}$ and self-energy $\Sigma$ (Eqs.~\ref{eq:ff-ch} and \ref{eq:chi_mat}). We have developed a novel algorithm based on a stochastic compression of the Lehmann representation of the Green's function~\cite{Altman2024-mp}, which significantly compresses the high-energy bands, and reduces the actual computational scaling with system size. The method amounts to modifying the KS energies $E_{n}$ and states $|\psi_n\rangle$ that are fed into the GW calculations. First, the energy spectrum of the KS states is partitioned into slices $\{S\}$ and a special protection region $P$ around the Fermi energy. The KS states and energies in $P$ remain untouched. The states in the remaining slices are replaced with stochastic linear combinations of the KS states within each slice, yielding stochastic pseudobands $    |\xi_j^S\rangle = \frac{1}{\sqrt{N_\xi}}\sum_{n\in S}e^{2\pi i \theta_n^j}|\psi_n\rangle$.
Here $\theta \in [0,1)$ is a random scalar, and we construct $N_\xi$ stochastic pseudobands for each slice, with $N_\xi$ typically between 2-5. We assign to these states $|\xi_j^S\rangle$ the average energy of the KS states in $S$.

The advantage of this approach is multiple-fold. First, because the slices are chosen according to their energy, they do not scale with system size. Second, through a careful error analysis, one can gradually increase the energy spanned by each slice, leading to an \textit{exponential} compression of the KS states necessary in the sums-over-bands in Eqs.~\ref{eq:ff-ch} and ~\ref{eq:chi_mat}.  
Finally, pseudobands capture the full KS Hamiltonian and eliminate band-truncation parameters in the calculation of $\chi$ and $\Sigma$.

To construct pseudobands, one needs to fully diagonalize the KS Hamiltonian, a bottleneck that scales as $O(N^3)$. We avoid this by rewriting pseudobands as random vectors $|x\rangle$ projected to the slice subspaces
$|\xi_j^S\rangle := f^S(H)|x\rangle$.
The projection operator $f^S(H) = \sum_{n\in S}|\psi_n\rangle\langle \psi_n|$ can be efficiently approximated using a Chebyshev-Jackson expansion $\tilde{f}^S_l(H)$ of order $l$~\cite{weisse2006kernel, schofield2012spectrum}.
In practice, the entire construction scales as a matrix-vector operation $\sim O(N) - O(N^2)$. 
The pseudobands and Chebyshev-Jackson methods alleviate a traditional bottleneck for sum-of-bands in GW calculations.

\subsection{Portability with Open Standard}\label{sec:portability}
Over the years, the optimization of BerkeleyGW~\cite{bgw} has enabled high-performance execution on leadership class HPC systems, from many-core CPU~\cite{DelBen2019a} to heterogeneous GPU architectures~\cite{DelBen2020_GB}, obtaining outstanding performance and achieving excellent time to solution. However, as HPC systems grow increasingly complex, with most computational power residing in specialized accelerators, it has been realized that a greater challenge lies in ensuring performance portability across HPC platforms. 

The BerkeleyGW's performance portability strategy is to leverage open, directive-based programming models, specifically OpenMP-target (OMP)~\cite{10.1007/978-3-030-85262-7_6} and OpenACC (OACC). 
Open standard programming models enable broad support across diverse architectures -- including NVIDIA, AMD, and Intel GPUs -- while preserving a unified codebase. This approach simplifies maintainability 
and helps obtain decent offloaded performance for novel developments. Despite the several advantages, we faced many challenges in  porting the pipeline as we navigated around compiler pitfalls, especially to: (i) support the programming model and stay up-to-date with the model standards, (ii) generate offloaded kernels ensuring correctness of results, (iii) interface with the vendor-specific libraries and corresponding APIs, and (iv) perform kernel optimizations capable of exploiting the characteristics of the specific hardware.

The BerkeleyGW GPU implementation demonstrates that open standards are not only practical for portability, but also capable of delivering high performance, with OpenACC recovering almost the entirety of the performance delivered by the best CUDA implementation on NVIDIA GPU~\cite{yang20208steps37tflops}. On the other hand, performance for the open models is lower on AMD and Intel GPUs, especially for large custom kernels not relying on vendors' offloaded libraries. The adopted portability strategy has been widely successful, with the public release of \textit{BerkeleyGW-4.0} (BGW-4.0)~\cite{doecode_124634} in production on NVIDIA, AMD and Intel GPU platforms alike, showcasing the effectiveness and readiness of open models in heterogeneous HPC environments moving forward.

\begin{table}[t!]
\caption{Application systems and computation sizes.}
\label{tab:parameters_for_systems}
\centering
\small
\setlength{\tabcolsep}{6pt}
\renewcommand{\arraystretch}{1.1}
\vspace{-0.3cm}
\begin{tabular}{c c c c c c}
\toprule
System Name & $N^{\psi}_G$ & $N_G$ & $N_b$ & $N_v$ & $N_c$ \\
\midrule
Si214    & 31,463  & 11,075  & $\gtrsim 5,500$  & 428  & $\gtrsim 5,000$ \\
Si510    & 74,653  & 26,529  & $\gtrsim 15,000$ & 1,020 & $\gtrsim 13,900$ \\
Si998    & 145,837 & 51,627  & $\gtrsim 28,000$ & 1,996 &$\gtrsim 26,000$ \\
Si2742   & 363,477 & 141,505 & 80,695 & 5,484 & 75,211 \\
Si2742'   & 363,477 & 141,505 & 15,840 & 5,484 & 10,356 \\
LiH998   & 81,313  & 52,923  & $\gtrsim 3,100$  & 499  &  $\gtrsim 2,600$ \\
LiH17574 & 506,991 & 362,733 & 49,920 & 8,787 & 41,133 \\
BN867 & 439,769 & 84,585 & 49,920 & 1,734 & 48,186 \\
\bottomrule
\end{tabular}
\end{table}

\subsection{Optimized GPP Kernel for Diagonal Self-Energy Matrix Elements} \label{sec:kernel_opt_diag}

\begin{figure}[t!]
  \centering
  \includegraphics[width=0.85\linewidth]{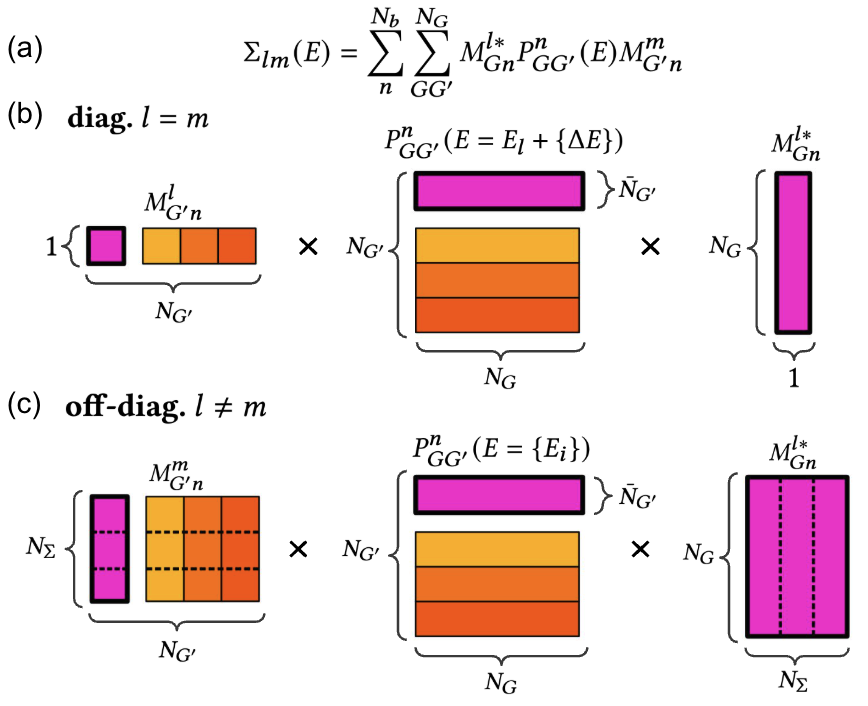} 
  \caption{Parallelization and data layout in Sigma-GPP. \textmd{ (a) GPP self-energy working equation. (b) Distribution of data in optimized GPP \textit{diag.} kernel. (c) Distribution of data in optimized GPP \textit{off-diag.} kernel. (b) and (c) represent one set of operations within the loop of summation over $n$. }
  \label{fig:DataLayout}}
  \vspace{3mm}
\end{figure}

The diagonal matrix elements of the self-energy operator directly provides information of quasiparticle energy levels, which are among the most commonly desired quantities.
The most computationally intensive GPP kernel (Fig.~\ref{fig:DataLayout}a) for the diagonal elements (denoted as \textit{diag.}) in the Sigma module is ported to the HIP and SYCL programming languages in order to gain the most optimal performance on Frontier and Aurora. CUDA version of the GPP \textit{diag.} kernel has been developed for NVIDIA GPUs~\cite{DelBen2020_GB}. In GPP \textit{diag.} kernel, we divide the computation over self-energy pools.  Within each pool, the GPP \textit{diag.} kernel computes its assigned self-energy matrix elements, e.g. $\Sigma_{ll}$, i.e., the diagonal ones. The GPP \textit{diag.} kernel is executed entirely on the device, with matrix elements $P^n_{GG'}$ being computed on the fly. The summation over all $N_{G'} (= N_G)$ is distributed over MPI ranks within a self-energy pool ($N_\text{rank}$ per pool) in the calculation, with each rank holding $\bar{N}_{G'} = N_{G'}/N_{\text{rank}}$ elements. Thus, in each kernel invocation, $N_G$ and $\bar{N}_{G'}$ are different bounds for the summation of $G$ and $G'$, respectively. The relation between loop indices $N_b < \bar{N}_{G'} \ll N_G$ dictates the design of the kernel (see Fig.~\ref{fig:DataLayout}b). In both HIP and SYCL kernels, we employ two levels of two-dimensional parallelism to decompose the problem. The first level decomposes the summation over $\bar{N}_{G'}$ and $N_b$  to distribute the computation over work-groups. The second level further decomposes the summation over $N_G$ within each work-group. This decomposition scheme effectively utilizes the accelerators' massive parallelism by maintaining high arithmetic intensity within each work-group.
In the following, we present detailed kernel adaptations to accelerators on Frontier and Aurora, where terms such as thread blocks (used in HIP and AMD architecture) and work-groups (used in SYCL and Intel architecture) are used interchangeably.


\subsubsection{Adapting to Frontier and Aurora Accelerators} \label{sec:kernel_opt_hip_sycl}
 The Frontier's AMD MI250X and Aurora's Intel PVC  GPUs have similar characteristics, therefore the GPP kernel optimizations share similar techniques on both architectures summarized below:
\begin{enumerate} [left=0pt]
    \item 
    Explicit memory management on device to coalesce memory access within a thread block. Prior to launching the kernel, one thread from each thread block loads sections of the $M_{Gn}^{l*}$ and $M_{G'n}^m$ arrays, corresponding to the current $\bar{N}_{G'}$ and $N_b$ blocks, into the shared memory. This significantly reduces the number of memory moves incurred during execution and drastically increases the arithmetic intensity. 
    \item Block size tuning of the second level of kernel parallelism 
    to maximize shared memory usage while avoiding memory overflow. We choose block sizes in the local parallelization to fully utilize Local Data Share (LDS) for each block on  GPU. 
    \item Loops are manually unrolled to gain maximum Vector General-Purpose Registers (VGPRs) and Scalar General Purpose Registers (SGPRs) occupancy without overflow. In particular, VGPRs overflowing on AMD accelerators incur huge latency in instruction execution. 
    By monitoring the VGPRs occupancy during compile time and carefully tuning loop unrolling and thread block sizes, we obtain over 97\% Vector Arithmetic Logic Unit (VALU) utilization rate while ensuring 0 VGPRs spillage.
    \item Replace expensive operations such as divisions and absolute values with reciprocals and multiplications as discussed in \cite{DelBen2020_GB}. This not only avoids the execution of such operations 
    but also adds to the Fused Multiply Add (FMA) instruction count, which utilizes hardware more efficiently. 
    In this way, the GPP kernel contains over 57\% FMA instructions with less than 4\% being inefficient transcendental operations.
    \item Two-stage reduction over thread blocks. First, each thread block designates a thread that accumulates the values using a masked intrinsic warp shuffle function. Then, we use the atomic add operation to accumulate the final result over all thread blocks. The choice of the number of thread blocks becomes a balance between register occupancy, memory access, and number of atomic reductions.
\end{enumerate}

\subsubsection{Hardware-Specific Adaptations} \label{sec:kernel_opt_hardware_specific}
The HIP and SYCL kernels have been adapted to better match the characteristics of each architecture to maximize hardware parallelization performance.
In the HIP kernel, we utilize a larger thread block size with more threads per block to maximize occupancy during execution. In the SYCL kernel, we instead tune the kernel to have more work-groups and fewer work-items per work-group to match the optimal Single Instruction Multiple Data (SIMD) width \cite{10.1145/3624062.3624177}. 
For AMD MI250X, more shared memory is loaded locally for each thread block with larger block size of computation to accommodate the increased memory. 
For Intel PVC, the layout of the work-groups require smaller chunks of shared memory for the large number of work-groups. These optimized memory layouts further decrease instruction stall rate and improve the arithmetic intensity.

\subsection{Optimized GPP Kernel for Off-Diagonal Self-Energy Matrix Elements}\label{sec:fullmat_kernel_opt}

Our optimized GPP \textit{diag.} kernel for diagonal matrix elements is at the ceiling of achievable arithmetic intensity considering its matrix-vector-like operation nature. Furthermore this implementation minimizes memory requirements by generating the band and frequency dependence of the inner matrix on the fly, which is highly efficient for diagonal-element calculations. On the other hand, advanced GW methods (including GWPT) require the calculation of full self-energy matrix including off-diagonal elements, thus the number of elements to be computed for $N_\Sigma$ bands becomes $N_\Sigma^2$ (i.e., the full matrix), instead of $N_\Sigma$ for diagonal-only elements. In this case we can gain arithmetic intensity by recasting the original GPP kernel into a matrix-matrix multiplication-like kernel. This is achieved by reformulating the formalism via generalizing the internal frequency argument $E$ in $\Sigma_{lm}(E)$ to a predefined uniform frequency grid $\{E_i\}$ independent of $(l, m)$ indices (in contrast to the GPP \textit{diag.} kernel) over the energy range of interest (e.g., the bandwidth across the $N_\Sigma$ bands). This generalization (see Fig.~\ref{fig:DataLayout}c) computes the full matrix of $\Sigma_{lm}(\{E\})$ with $l$ and $m$ span $N_\Sigma$, offering much more accurate self-consistent quasiparticle energies from the full solutions of the Dyson’s equation and dynamical behavior of the electron-phonon matrix elements from GWPT.

We have implemented a new full-matrix GPP kernel which efficiently computes off-diagonal matrix elements (denoted as \textit{off-diag.}).
To increase arithmetic intensity, in the GPP \textit{off-diag.} kernel, we precompute the band ($n = 1, ..., N_b$) and frequency ($\{E_i\}_{i = 1, ..., N_E}$) dependent matrices $P$ (Fig.~\ref{fig:DataLayout}c) over all $(n,E)$ pairs, and perform ZGEMM for each $(n,E)$ configuration.
The pre-computation (\textit{prep.} step) utilizes the same optimizations as in the GPP \textit{diag.} kernel (Sec. \ref{sec:kernel_opt_hip_sycl} and \ref{sec:kernel_opt_hardware_specific}). For diagonal-only calculations, this new strategy provides no benefit, because it significantly increases the memory demands, and the overhead of the \textit{prep.} step cancels the performance gain from ZGEMM. However, when the full $\Sigma(E)$ matrix is required for large $N_\Sigma$ (for full solutions of GW Dyson's equation and large-scale GWPT calculations), the reuse of the precomputed $P$ matrices for many target states makes this new formulation very competitive and efficient. The resulting computation of the GPP \textit{off-diag.} kernel reduces to two consecutive dense matrix multiplications (ZGEMM) of dimensions $N_\Sigma \times N_G \times N_G$ and $N_\Sigma \times N_G \times N_\Sigma$ per $(n,E)$ pair (Fig.~\ref{fig:DataLayout}c), achieving a two-fold increase in performance throughput compared to the GPP \textit{diag.} kernel.

\section{How Performance Was Measured}


The performance measurements are obtained on a set of realistic applications (see Table \ref{tab:parameters_for_systems}) to highlight the capabilities of BerkeleyGW across methodologies and system sizes. 
Our general baseline performance
features semiconductor defects (proxy for solid-state qubits) with systems of variable sizes from small (214 silicon atoms) to large (2742 silicon atoms)~\cite{N10_benchmark}. 
The largest system Si2742 contains a total of 80,695 bands. The mixed stochastic-deterministic pseudobands approach allows for improved convergence at a lower number of band, i.e., at $N_b = 15,840$. We label the same system with 15,840 bands as Si2742'.
We also present massive applications of GW calculations on defects in solid LiH, with supercells containing up to 17,574 atoms, surpassing the previously reported  largest GW calculation of 13,824-atom LiH~\cite{10.1109/SC41406.2024.00067}. 
Additionally, we report results of 3.88$^\circ$-twisted BN moir\'e bilayer consisting of 867 atoms (with 1.5-nm vacuum layer), with a carbon substitution at a boron site adjacent to a nitrogen vacancy. 
Defects in layered BN are useful as single-photon emitters, with moir\'e twisting offering tunability.
To demonstrate the electron-phonon coupling capabilities, we perform GWPT calculations on a LiH defect system with 998 atoms, involving six atomic displacements ($N_p = 6$). These calculations help describe quantum decoherence and excitation lifetimes.

The performance results are collected from three HPC systems:
\begin{itemize} [left=0pt]
     \item \emph{Frontier} (OLCF): 9,408 nodes, each with 1 AMD Milan CPU and 4 AMD Instinct MI250X GPUs, each comprised of 2 Graphics Compute Dies (GCD) for a total of 8 devices. FP64 peak performance per GPU 23.9 TeraFLOP/s and aggregated 1.80 ExaFLOP/s.
     \item \emph{Aurora} (ALCF): 10,624 nodes, each with 2 Intel Xeon CPU Max Series and 6 Intel Data Center GPU Max Series ``Ponte Vecchio'' (PVC), each comprised of 2 tiles for a total of 12 devices. FP64 peak per GPU 17 TeraFLOP/s and aggregated 2.17 ExaFLOP/s. \textbf{Note:} At the time of this work, Aurora's GPUs are not running at full capacity. Therefore, we compare against the measured FP64 Vector MAD Peak of 11.4 TeraFLOP/s using Intel Advisor Profiler. Hence, the attainable peak of Aurora is 1.45 ExaFLOP/s.
     \item \emph{Perlmutter} (NERSC):  1,792 nodes, each with 1 AMD Milan CPU and 4 NVIDIA A100 GPUs. FP64 peak performance per GPU 9.7 TeraFLOP/s and aggregated 69.5 PetaFLOP/s.
\end{itemize}
Unless otherwise stated, in this work, a ``GPU'' means a single NVIDIA A100 for Perlmutter, a single MI250X's GCD for Frontier, and a single PVC's tile for Aurora.

\begin{table}[t!]
\caption{FLOP count from measured (Meas.) and estimated (Est.) performance for Si-214 on Frontier (F) and Aurora (A). }
\label{tab:Si214_frontier_aurora}
\centering
\small
\setlength{\tabcolsep}{4pt}
\renewcommand{\arraystretch}{1.1}
\vspace{-0.3cm}
\begin{tabular}{c c c c c c c c}
\toprule
& $N_\Sigma$ & $N_b$ & $N_G$ & $N_E$ & \shortstack{Est.\\(TFLOP)} & \shortstack{Meas.\\(TFLOP)} & Accuracy \\
\midrule
\multirow{3}{*}{F}   & 2  & 5,000  & 3,911  & 3 & 38.32     & 38.55     & 99.39\% \\
                     & 4  & 15,045 & 26,529 & 3 & 10,609.67 & 10,564.75 & 99.57\% \\
                     & 8  & 6,340  & 11,075 & 4 & 2,077.88  & 2,064.84  & 99.37\%\\
\midrule
\multirow{3}{*}{A}& 2 & 3,000 & 11,075 & 6 & 416.27 & 415.17 & 99.74\% \\
                  & 1 & 5,000 & 11,075 & 6 & 346.89 & 345.89 & 99.71\% \\
                  & 1 & 2,000 & 11,075 & 6 & 138.76 & 139.42 & 99.52\% \\
\bottomrule
\end{tabular}
\end{table}

To determine the number of floating point operations (FLOPs) performed in the Sigma module, we use the canonical FLOP count from the most computationally intensive kernel.
In the Sigma module, the GPP kernel takes up over 95\% of the FLOPs for production calculations. 
The computational complexity for the GPP \textit{diag.} kernel is $O(N_\Sigma N_b N_G^2 N_E)$.  Through a series of tests listed in Table. \ref{tab:Si214_frontier_aurora}, we determine a linear relationship between the FLOP count and the computational complexity as,
\begin{equation} \label{eq:flop_count}
    \text{FLOP count (GPP \textit{diag.})} = \alpha \times N_\Sigma N_b N_G^2 N_E \ \ ,
\end{equation}
where $\alpha$ is an architecture- and compiler-dependent constant prefactor. To determine the quasiparticle energy $E_i^\text{QP}$ from the self-consistent relation in Eq. \ref{eq:dyson_equation}, we need a few $N_E \sim O(1) - O(10)$ sampling points for $E$ in evaluating specific diagonal element $\Sigma_{ll}(E)$, where the value of $E$ depends on the $l$ index. 
We use the ROCm profiler for AMD GPUs on Frontier and the Intel Advisor profiler for Intel GPUs on Aurora to determine the prefactor value $\alpha$. 
The prefactor for GPP \textit{diag.} kernel on Frontier and Aurora are measured to be $\alpha_\text{Frontier} = 83.50$ and $\alpha_\text{Aurora} = 94.27$, respectively.
In Table \ref{tab:Si214_frontier_aurora}, we also verify the accuracy of the prefactor with less than 1\% discrepancy between the estimated and measured FLOP count.

For the GPP \textit{off-diag.} kernel, we account only for the ZGEMM operations, with the \textit{prep.} step contributing diminishing fraction of FLOPs at large $N_\Sigma$ and $N_E$. We sample uniformly $N_E \sim O(10^2)$ points for $E$ in $\Sigma_{lm}(E)$, reformulated to be independent of $(l, m)$ to cast the algorithm to ZGEMM, spanning the energy window of $N_\Sigma$ bands for full solutions of Dyson's equation.  Our implementation performs $2N_b N_E$ times of ZGEMM operations, with the number of FLOPs counted as (based on standard ZGEMM FLOP count):
\begin{equation} \label{eq:flop_count_fullmat}
\begin{split}
    & \text{FLOP count (GPP \textit{off-diag.}; ZGEMM only)}  \\
    = \  &  2 N_b N_E \times 8  (N_\Sigma N_G^2 + N_G N_\Sigma^2) \ \ .
\end{split}
\end{equation}
Note that in the GPP \textit{off-diag.} kernel, we only count FLOPs from ZGEMM, but still measure the full kernel runtime (including \textit{prep.} step), hence our reported performance stands as a lower bound.

\begin{table*}[ht!]
 \caption{Sigma time to solution (seconds) for Si-510 with $N_\Sigma=128$ across architectures and programming models.}
\label{tab:Si510_time_to_solution}
\centering
\small
\vspace{-0.3cm}
\begin{tabular}{ c c c c c c c c c c c c c c}
 \toprule
 & \multicolumn{10}{@{}c}{GW-GPP \textit{diag.}} & \multicolumn{3}{@{}c}{GW-FF} \tabularnewline
   \cmidrule(lr){2-11} \cmidrule(lr){12-14} 
 & \multicolumn{4}{@{}c}{Perlmutter} & \multicolumn{3}{@{}c}{Frontier} & \multicolumn{3}{@{}c}{Aurora} & 
   \multicolumn{1}{@{}c}{Perlmutter} & \multicolumn{1}{@{}c}{Frontier} & \multicolumn{1}{@{}c}{Aurora} \tabularnewline
 \cmidrule(lr){2-5}\cmidrule(lr){6-8}\cmidrule(lr){9-11}\cmidrule(lr){12-12}\cmidrule(lr){13-13}\cmidrule(lr){14-14}
\# of Nodes  &  OMP$^\dagger$ & OMP & OACC    & CUDA    &  OMP$^\dagger$   &  OACC  &     HIP &     OMP$^\dagger$ & OMP &   SYCL & OACC & OACC & OMP \tabularnewline 
 \midrule
   4     & 4,186.3 &  3,268.7 & 3,197.3 & 2,928.3  & 2,562.1  & 2,111.9 &  1,382.5 &  3,621.1 &  2,877.2  & 1,416.0 & 528.2 & 354.4 &  364.7 \tabularnewline
   8     & 1,978.9 &  1,640.2 & 1,601.1 & 1,467.1  & 1,294.9  & 1,062.7 &   684.6 &  1,835.2 &  1,437.9  &  736.0 & 281.8 & 188.3 &  208.3 \tabularnewline
  16     &  990.1 &   826.0 &  804.6 &  744.2  &  654.9  &  548.6 &   369.3 &   918.5 &   727.1  &  390.0 & 159.3 & 112.7 &  128.2 \tabularnewline
  32     &  501.9 &   419.7 &  407.8 &  383.8  &  336.8  &  282.0 &   191.4 &   467.6 &   372.6  &  205.3 &  99.22 &  70.6 &   93.9 \tabularnewline
  64     &  260.1 &   218.3 &  214.7 &  203.5  &  182.7  &  147.3 &   110.5 &   245.6 &   199.1  &  121.6 &  71.5 &  53.7 &   69.9 \tabularnewline
 \bottomrule
\end{tabular}
\end{table*}

\section{Performance Results}

\subsection{Performance Portability}

We evaluate the performance portability of several GW implementations across different programming models and hardware architectures, as summarized in Table~\ref{tab:Si510_time_to_solution}. We focus on five programming models: two directive-based open standards (OpenMP and OpenACC) and three hardware-optimized models (CUDA, HIP, and SYCL). These implementations are benchmarked on different GPU architectures, namely NVIDIA, AMD, and Intel.

\begin{figure}[b!]
  \centering
  \includegraphics[width=0.8\linewidth]{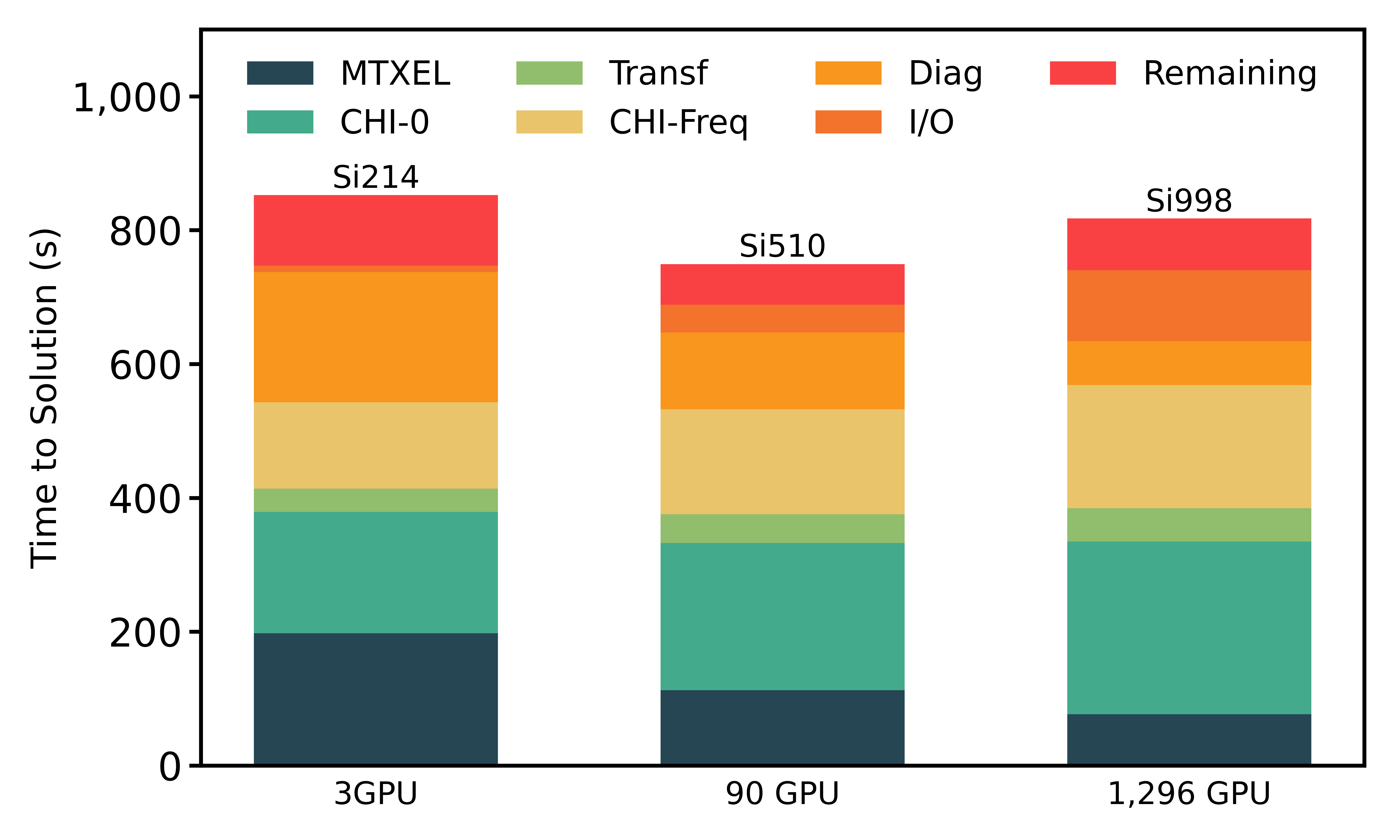} 
  \caption{Weak scaling of the GW-FF Epsilon on Aurora.
  \label{fig:FF_weak}}
\end{figure}

Open standards such as OpenACC and OpenMP remain valuable tools for achieving performance portability, particularly as compiler technology continues to mature. On NVIDIA hardware, OpenACC demonstrates exceptional performance, recovering over 90\% of the best CUDA implementation, highlighting good compiler support. However, the situation is less favorable on Frontier (AMD GPUs), where OpenACC gives only 60--70\% of our best HIP performance, which is likely attributed to overall less maturity and less aggressive compiler optimization capabilities. Furthermore, OpenACC is not currently supported by Intel compilers on Intel GPUs, limiting its portability across all major vendors. The OpenACC implementation discussed here is publicly available as part of the released BGW-4.0. 

For OpenMP, two implementations have been evaluated here. The first one, labeled as OMP$^\dagger$ in Table \ref{tab:Si510_time_to_solution}, was also released in BGW-4.0, and at the time of release, it did not incorporate all the optimizations present in the OpenACC version. As a result, OMP$^\dagger$ is approximately 15--20\% slower than OpenACC on both Perlmutter and Frontier. The second OpenMP version, labeled as OMP, includes additional optimizations similar to our OpenACC implementation, including reduced kernel branching and improved data reuse. However, it still lacks support for asynchronous GPU execution. Despite this, it nearly matches the performance of OpenACC on Perlmutter. On Frontier, however, this optimized OMP implementation performs poorly, taking long execution time even for small applications. The issue appears to arise from the compiler's attempt to parallelize over the innermost strided loops of the kernel, which are correctly serialized in the OpenACC version using \textit{loop seq}. 
On Intel GPUs, the optimized OpenMP version provides around 20\% better performance than OMP$^\dagger$, but is still 50\% slower than the SYCL implementation. These results underscore current limitations of the OpenMP kernel optimization capabilities on Intel software and hardware.

While these results may suggest an insurmountable performance advantage for hardware-optimized programming models, especially on non-NVIDIA hardware, we emphasize that our goal is to evaluate how current implementations perform \textit{out of the box}. 
We anticipate that, with continued improvements in compilers and tool chains, the open standards could catch up to their hardware-optimized counterparts across all major GPU architectures. Despite current limitations, the results presented here are encouraging, showing that open standards already deliver competitive performance across different hardware platforms, 
and hold promise as viable, maintainable, and portable solutions for future architectures.

\begin{figure}[b!]
  \centering
  \includegraphics[width=1.0\linewidth]{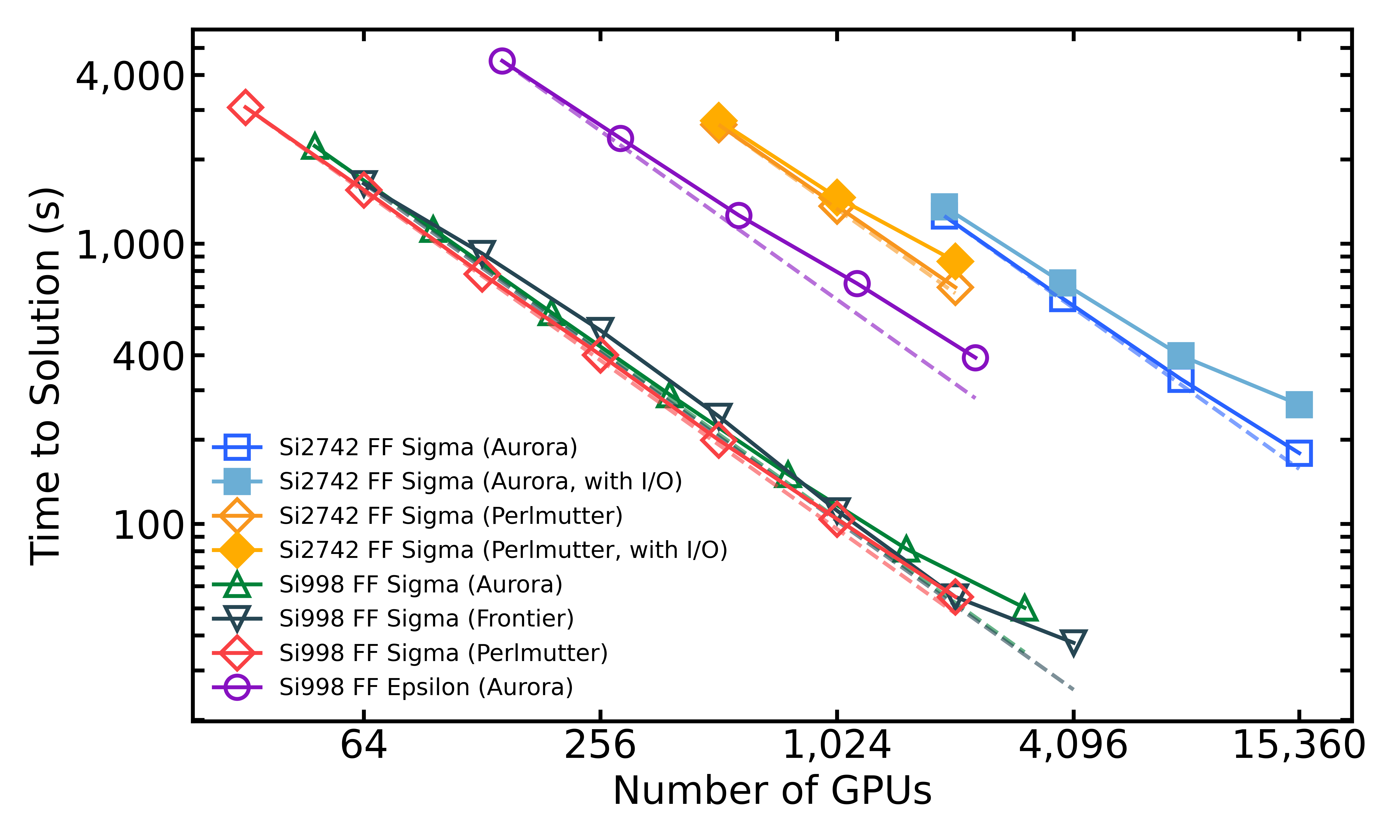} \Description{A} \\ \Description{B}
  \caption{Strong scaling of the GW-FF. \textmd{ Results are reported excluding I/O unless noted specifically.}
  \label{fig:FF_strong}}
\end{figure}

\subsection{Performance of GW-FF}

The full-frequency GW implementation in BerkeleyGW is only slightly more costly than the GPP method, due to the use of the static subspace approximation. 
This implementation is highly scalable due to the multi-layer parallelizations (including the additional level over frequencies), showing strong scaling up to thousands of GPUs, with portability across all three major vendors.

\begin{figure}[t!]
  \centering
  \includegraphics[width=0.9\linewidth]{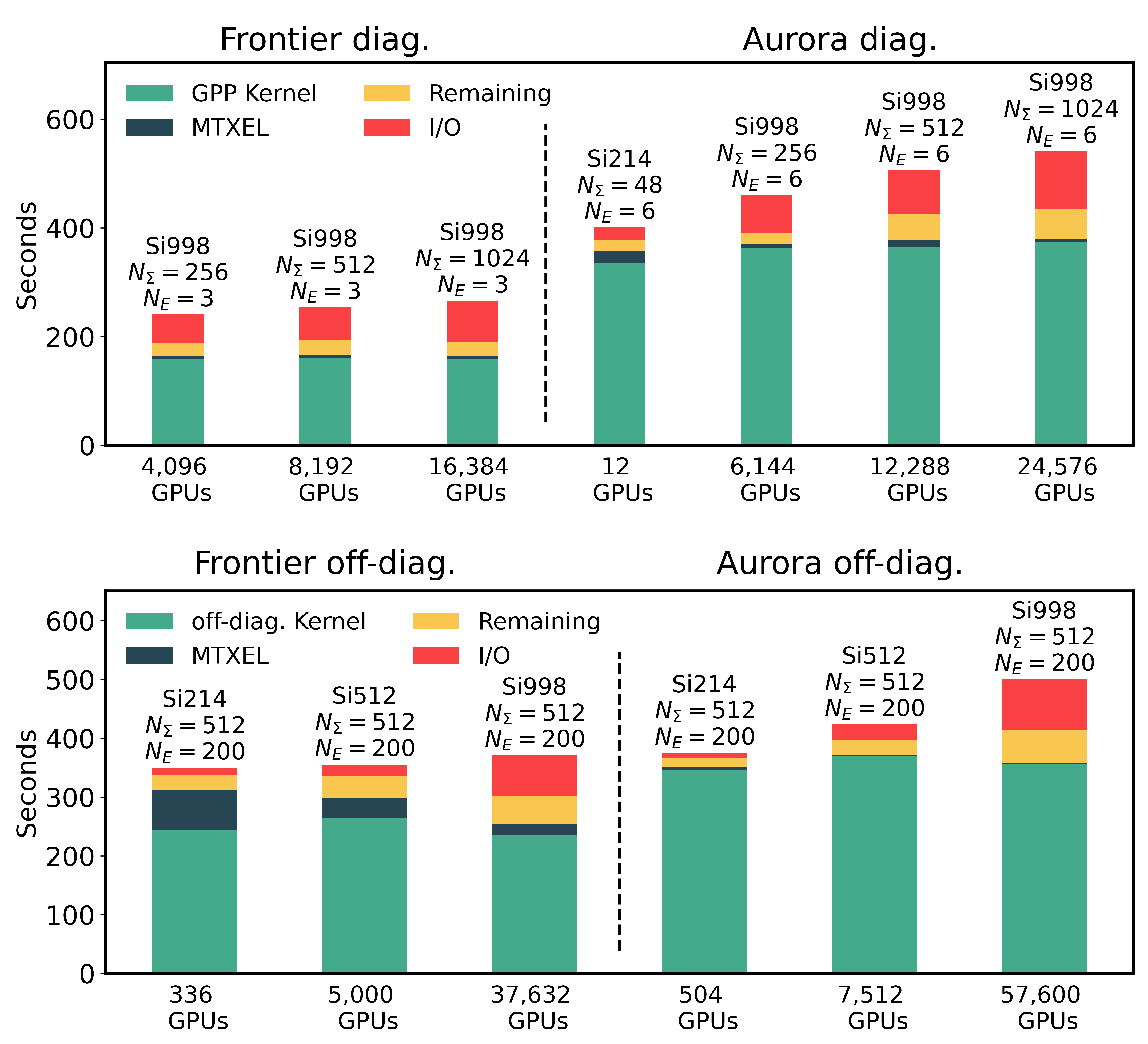}
  \caption{Weak scaling of the GW-GPP Sigma.}
  \label{fig:weak_scaling}
  \vspace{5mm}
\end{figure}

In the Epsilon module, the computational cost for full-frequency polarizability is only about twice as high as for the GPP model. 
The weak scaling of the FF implementation is shown in Fig.~\ref{fig:FF_weak}. The main computational kernels (\textit{CHI-0}, \textit{CHI-Freq}, and \textit{Transf}) show nearly ideal weak scaling, while the lower scaling kernels (\textit{MTXEL} and \textit{Diag}) decrease significantly. 
In this case, the additional calculation of 19 frequencies with $\sim20\%$ subspace fraction only takes about the same time as the initial zero-frequency calculation with the full planewave basis. 

The calculation of self-energy in Sigma using the full-frequency polarizability becomes very efficient with the static subspace approximation. Furthermore, the extreme parallelism offered by the number of self-energy elements allows for strong scaling up to tens of thousands of GPUs and portable scaling on all three HPC systems, as shown in Fig.\ref{fig:FF_strong}. 
Weak scaling with respect to $N_\Sigma$ shows the same favorable performance up to tens of thousands of GPUs as the GPP model, since the parallelization scheme is identical. Weak scaling with the compute pool size is less favorable due to communication, but the abundant parallelism available over $N_\Sigma$ alleviates this issue in large-scale calculations.





\subsection{Optimized GPP Kernel Performance}

Using the hardware-optimized programming models, we performed systematic scaling calculations to demonstrate the performance of the GPP implementations. We report results on Frontier and Aurora, where the kernel implementations are optimized with HIP and SYCL, respectively.

Fig. \ref{fig:weak_scaling} shows weak scaling on both Frontier and Aurora with varying system sizes. The problem size is scaled based on Eqs.~\ref{eq:flop_count} and~\ref{eq:flop_count_fullmat}. The dominant computational step, the GPP \textit{diag.} and \textit{off-diag.} kernels, construct the GW self-energy operator and its matrix elements using the GPP model. We observe excellent weak scaling and time to solution up to tens of thousands GPUs on both Frontier and Aurora systems.

\begin{figure}[t!]
  \centering
  \includegraphics[width=0.95\linewidth]{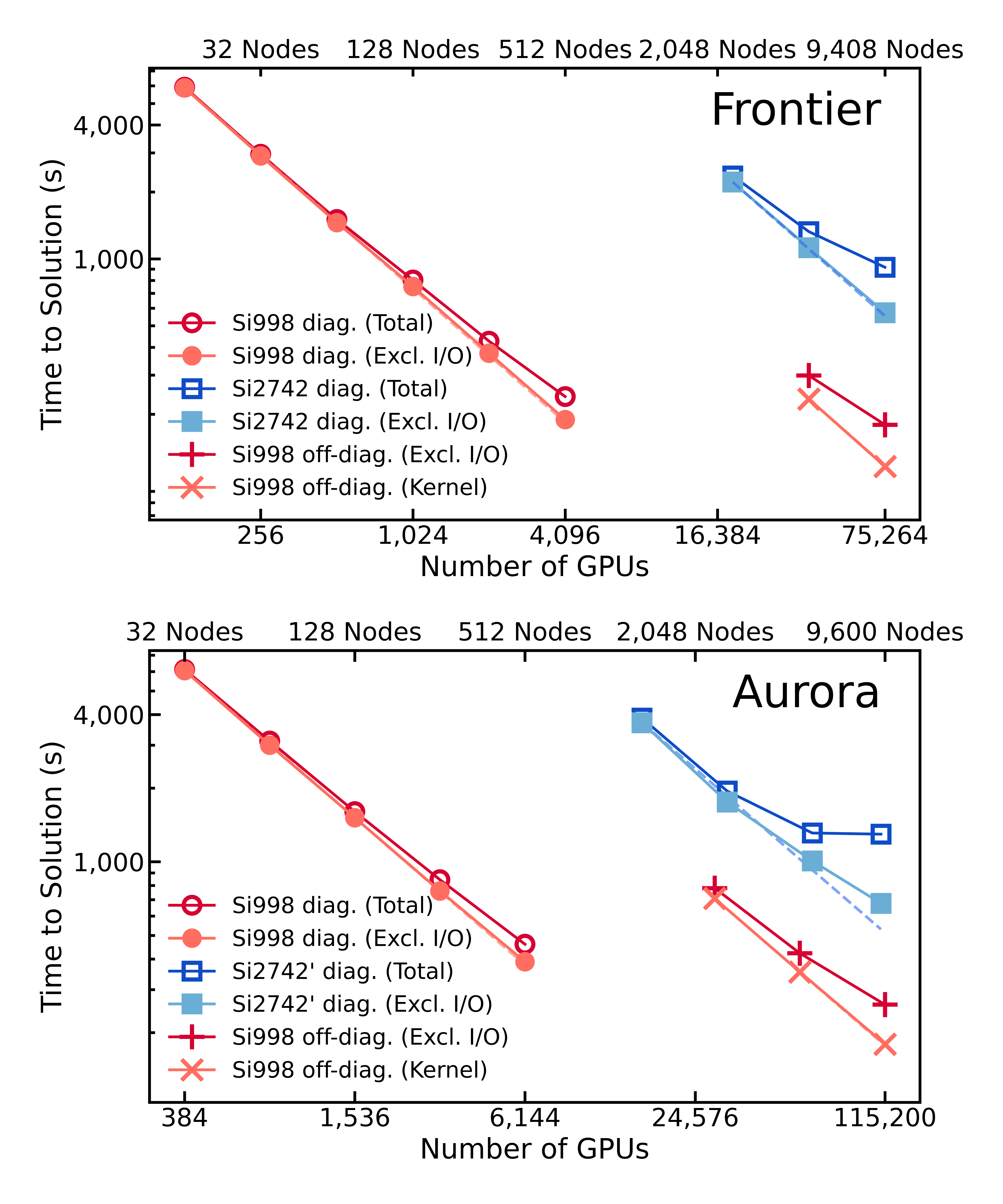}
  \caption{Strong scaling of the GW-GPP Sigma. 
  }
  \label{fig:combined_strong}
  \vspace{5mm}
\end{figure}

Fig. \ref{fig:combined_strong} shows strong scaling of GPP \textit{diag.} and \textit{off-diag.} calculations on Frontier and Aurora using Si998 and Si2742 systems. Our results show excellent strong scaling excluding I/O, up to the full machine of Frontier with 9,408 nodes, and up to 90.4\% of the full machine of Aurora with 9,600 nodes. 
The GPP \textit{diag.} kernel shows excellent scalability across a small to large number of GPUs due to its memory-efficient formalism and implementation.
The GPP \textit{off-diag.} kernel shows the best kernel performance with large-scale calculations, by leveraging the ZGEMM library, and particularly the matrix cores of AMD GPUs on Frontier.
Moreover, we have explored the Tensile library on Frontier, which optimizes ZGEMM performance for specific matrix sizes in the GPP \textit{off-diag.} kernel. Our observations show that for the large application case (Si998 with $N_\Sigma = 512$), the default ZGEMM library call already reaches the best-achievable performance compared to the one with Tensile optimization, whereas for moderate problem size (Si998 with $N_\Sigma = 384$), the Tensile optimization can boost the overall kernel performance by $\sim 10\%$, to the similar best achievable level.

\subsection{Full System Runs and Peak Performance}

\begin{figure}[t!]
  \centering
  \includegraphics[width=0.95\linewidth]{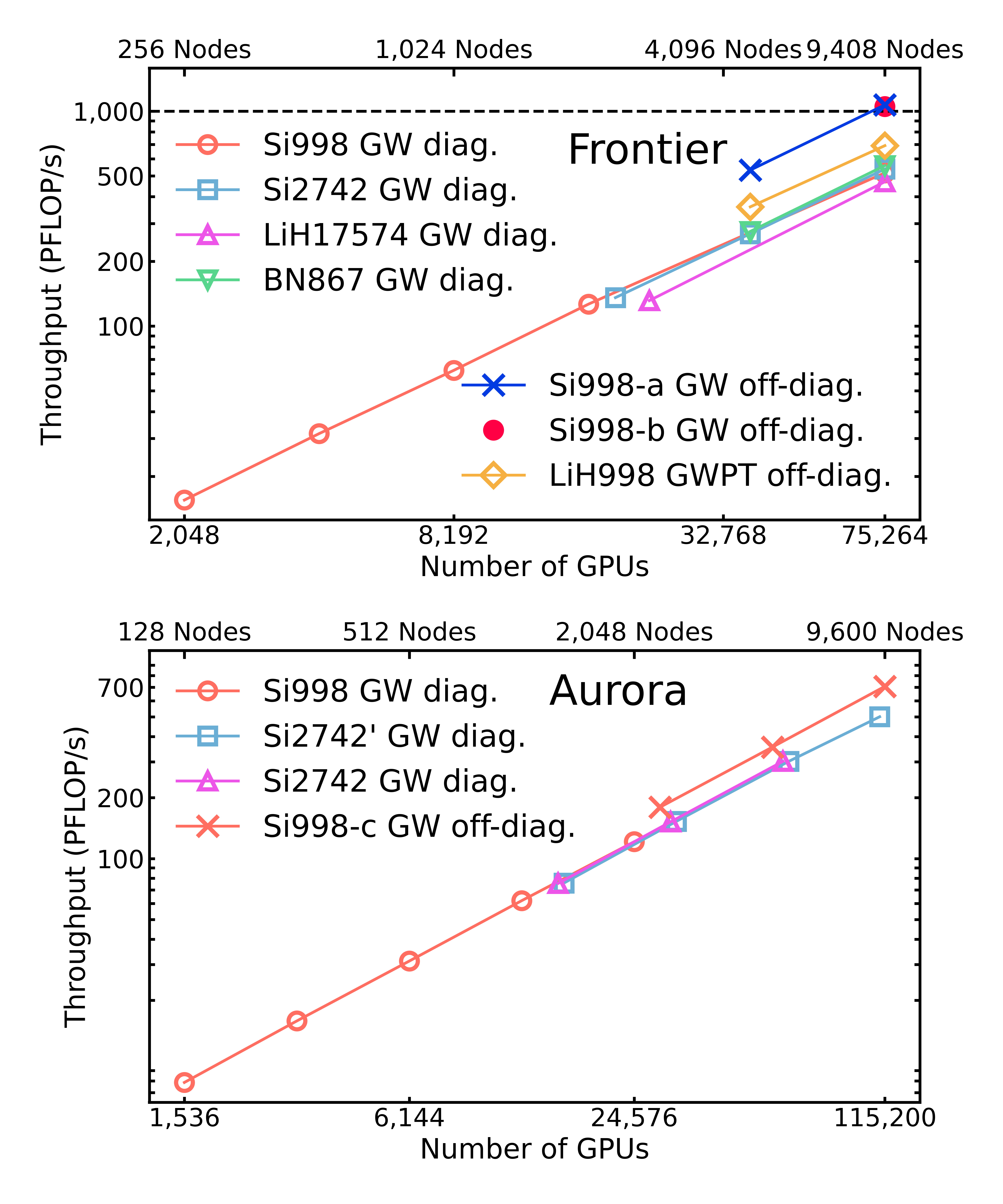}
  \caption{Throughput of GPP kernel achieved on Frontier and Aurora. \textmd{ The dashed line (upper panel) marks 1.0 ExaFLOP/s performance. Si998 demonstrates multiple configurations: Si998-a ($N_E = 200, N_b=28,224$), Si998-b ($N_E = 512, N_b = 28,224$), and Si998-c ($N_E = 200, N_b = 28,800$).} }
  \label{fig:combined_throughtput}
\end{figure}

Fig.~\ref{fig:combined_throughtput} shows the throughput performance of the Sigma-GPP \textit{diag.} and \textit{off-diag.} kernels (the most computationally intensive kernels) on Frontier and Aurora. Here, we demonstrate applications over a wide range of systems, including solid-state defects of Si and LiH, and defects in BN moir\'e superlattices.

On both Frontier and Aurora, the GPP \textit{diag.} kernel consistently reaches $\sim500$ PetaFLOP/s at (nearly) the full machine scale with the hardware-optimized implementations. In particular, at full machine of Frontier (9408 nodes, or 75,264 AMD GPUs), we achieved 558.3 PetaFLOP/s in double precision, corresponding to 31.04\% of the theoretical peak; and with 87.5\% of the full Aurora (9,296 nodes, or 111,552 Intel GPUs), we achieved 500.97 PetaFLOPs in double precision, corresponding to 39.39\% of the attainable peak. 

The GPP \textit{off-diag.} kernel at (nearly) the full machine scale achieves significantly higher double-precision throughput performance: \textbf{1.069 ExaFLOP/s on 9,408 Frontier nodes} (75,264 AMD GPUs, full machine), corresponding to \textbf{59.45\% of the theoretical peak}; and \textbf{707.52 PetaFLOP/s on 9,600 Aurora nodes} (115,200 Intel GPUs, 90.4\% of full machine), corresponding to \textbf{48.79\% of attainable peak}. These performance gains are directly related to the reformulation of the most computationally heavy contractions into ZGEMM operations, benefiting the off-diagonal calculations. This recasting substantially increases arithmetic intensity at the cost of additional memory consumption. The resulting trade-off between memory footprint and compute efficiency proves highly favorable when a large number of $N_\Sigma$ is calculated along with a fine grid of $N_E$. 

In Table~\ref{tab:best_throughput}, we list some of the best achieved throughput results.
The excellent scalability up to over 1.0 ExaFLOP/s and high peak percentage ($\sim50 - 60 \%$) of the theoretical or attainable peak in \textit{double precision} have clearly established the effectiveness and generalizability of our kernel optimizations across major GPU architectures by different vendors. 
Note that the performance of the whole application improves with the desired accuracy. For instance, in the Si998-b case (Table~\ref{tab:best_throughput}), computing $N_E = 512$ frequencies yields a kernel performance of 1.051 ExaFLOP/s, along with excellent whole application performance of over 800 PetaFLOP/s excluding I/O, and over 500 PetaFLOP/s including I/O.
Our work not only demonstrates flexibility of directive-based open programming models, but also highlights the transferable knowledge in hardware-optimized implementations for achieving high peak performance.

\section{Implications}

This work presents several key innovations of BerkeleyGW benchmarked on the exascale Aurora and Frontier supercomputers. On the HPC side, we have successfully enabled true portability using both directive-based open standards and hardware-optimized models, achieving high performance on AMD, Intel, and NVIDIA GPU architectures.  Specifically, we have scaled the GW calculations to (nearly) the full machine of Frontier and Aurora, obtained $\sim700$ to over 1,000 FP64 PetaFLOP/s kernel performance ($\sim50-60\%$ of the theoretical/attainable peak). 
On the methodology aspect, we have successfully enabled large-scale and highly efficient GWPT calculations for correlated electron-phonon coupling, along with scalable and portable GW calculations using both GPP and full-frequency schemes. 
The mixed stochastic-deterministic pseudobands method can further help to reduce the scaling of the GW methods. These versatile functionalities place BerkeleyGW at the forefront of first-principles many-body perturbation theory research.

With these advancements, BerkeleyGW is now able to systematically compute at scale the quasiparticle excited-state properties and the electron-phonon coupling phenomena, which are critical to transport, optical absorption, and decoherence and lifetimes of quantum states (e.g. in qubits and quantum emitters), for complex materials structures with  $O(10^3) - O(10^4)$ atoms.
The portable and efficient utilization of resources, along with the capability to describe heterogeneous systems and quantum many-body interactions, sets up a new frontier for studying increasingly complex quantum materials, phenomena, and devices in the exascale age.

\begin{table}[t!]
\caption{Best throughput performance on Frontier (F) and Aurora (A).}
\label{tab:best_throughput}
\centering
\small
\setlength{\tabcolsep}{4pt}
\renewcommand{\arraystretch}{1.1}
\vspace{-0.3cm}
\begin{tabular}{c c c c c c}
\toprule
\multicolumn{6}{c}{\textbf{Optimized diagonal GPP Kernel}} \\
\midrule
System & Calculation & \shortstack{\# of \\Nodes} & \shortstack{Time\\(s)} & \shortstack{Perf.\\(PFLOP/s)} & \shortstack{\% of\\Peak} \\
\midrule
BN867 GW & Kernel (F) & 9,408 & 188.45 & 558.32 & 31.04 \\
Si2742 GW &  Kernel (F)  & 9,408 & 445.02 & 534.80 & 29.73 \\
Si2742' GW &  Kernel (A)    & 9,296 & 475.58 & 500.97 & 39.39 \\
LiH998 GWPT &  Kernel (F)  & 9,408 & 92.91 & 479.27 & 26.64 \\
\midrule
\multicolumn{6}{c}{\textbf{Optimized off-diagonal GPP Kernel}} \\
\midrule
System & Calculation & \shortstack{\# of \\Nodes} & \shortstack{Time\\(s)} & \shortstack{Perf.\\(PFLOP/s)} & \shortstack{\% of\\Peak} \\
\midrule
Si998-a GW & Kernel (\textbf{F}) & 9,408 & 116.4 & \textbf{1,069.36} & \textbf{59.45} \\
Si998-b GW & Kernel (F) & 9,408 & 303.13 & 1,051.21 & 58.44 \\
Si998-b GW &  Tot. excl. I/O (F)  & 9,408 & 390.75 & 815.49 & 45.33 \\
Si998-b GW &  Tot. incl. I/O (F)  & 9,408 & 604.96 & 526.73 & 29.28 \\
Si998-c GW & Kernel (\textbf{A}) & 9,600 & 179.52 & \textbf{707.52} & \textbf{48.79} \\
LiH998 GWPT &  Kernel (F)  & 9,408 & 30.13 & 691.10 & 38.42 \\
\bottomrule
\end{tabular}
\end{table}

\begin{acks}
This work was primarily supported by C2SEPEM at LBNL, which is funded by the US Department of Energy (DOE) under contract No. DEAC02-05CH11231, which supported the core developments of advanced theories, methods, algorithms and exascale optimizations in BerkeleyGW (B.Z., A.R.A., Y.S., S.G.L., J.R.D., F.H.J., Z.L., and M.D.B.). 
B.Z., C.E.H. and Z.L. acknowledge US NSF (DMR-2440763) for partial support for numerical calculations.
A.R.A. acknowledges the US DOE Center NPNEQ (DE-AC52-07NA27344) for Parabands development.
Y.S. acknowledges NSF (DMR-2238328) for Chebyshev-Jackson algorithm development.
D.V.F. and D.W. acknowledge US DOE BEAST project (DE-SC0022247) for the GW-FF offloaded implementation.
An award of computer time was provided by the INCITE program (\textit{MAT280, ElectronPhonon}) and by ALCC (\textit{CPH169, PredictPhotocatal}) program. This research used resources of both the ALCF and OLCF which are DOE SC User Facilities supported under contracts DE-AC02-06CH11357 and DE-AC05-00OR22725.
Computational resources were also provided by DOE User Facility NERSC (DE-AC0205CH11231), and by TACC (\textit{DMR22042}), which is supported by NSF (OAC-1818253).
We thank Dr. Ye Luo at ALCF for helpful discussions.
\end{acks}


\bibliographystyle{ACM-Reference-Format}
\bibliography{references}  


\end{document}